# A Physiologically-based simulation model of color appearance for red-green color vision deficiency

Lijia Sun,[1] Shining Ma,[1,*] Yong Tao,[2] Liang Jia,[3] Yue Liu,[1] Yongtian Wang,[1] and Weitao Song[1,4]

[1]Beijing Engineering Research Center of Mixed Reality and Advanced Display, School of Optics and Photonics, Beijing Institute of Technology, Beijing 100081, China
[2]Department of Ophthalmology, Beijing Chaoyang Hospital, Capital Medical University, Beijing 100020, China
[3]Senior Department of Ophthalmology, the Third Medical Center, Chinese PLA General Hospital, Beijing 100039, China
[4]swt@bit.edu.cn
*shining.ma@bit.edu.cn

**Abstract:** Various simulation methods of color appearance for dichromats or anomalous trichromats have been proposed over the years. To further improve the performance of the simulation model and extend the application range to both dichromats or anomalous trichromats, we have proposed a simulation model of cone fundamentals specifically designed for individuals with red-green type color vision deficiency (CVD) based on the CIE 2006 physiological observer model. By utilizing the simulated cone fundamentals, it becomes possible to predict the color appearance of real scenes and digital images for CVD. The fundamental premise of the new model is rooted in the hypothesis that CVD arises from a shift in the peak wavelength of the photopigment absorption spectrum of the L or M cone. Instead of simply maintaining the waveform without alteration as observed in prior studies, we altered waveforms of the absorption spectra of anomalous L/M cone photopigments when adjusting their peak wavelengths. Regarding different shapes in the absorption spectrum between the L and M cone, the absorption spectrum of the anomalous L/M cone was obtained by combining the peak wavenumber shift and linear interpolation of spectral quantal absorption curves between L- and M-photopigments in the wavenumber domain. The performance of the proposed model was substantiated through experimental validation by the pseudoisochromatic plates and Farnsworth Munsell 100 Hue test (FM-100). The findings revealed a high level of consistency between the model prediction and the actual perception reported by individuals with CVD.

## 1. Introduction

Color vision deficiency (CVD) affects up to 8% of males and 0.5% of females [1], impairing individuals in effectively conducting color-related visual tasks. Observers with color vision deficiency (CVD) exhibit an anomalous absorptance spectrum of photopigment for any of the three cones. Substantial evidence indicates that the abnormality in color vision results from a shift in the peak wavelength of the absorptance spectrum for the anomalous photopigment, a phenomenon physiologically determined by the type of photopigment opsins [2]. In cases of anomalous trichromacy, the alteration in the absorptance spectrum depends on the type of photopigments, with anomalous L-, M-, S-cones corresponding to protanomaly, deuteranomaly, tritanomaly, respectively. In the cases of dichromacy, one photopigment is either absent or replaced by another type, with the absence of L-, M-, or S-cones corresponding to protanopia, deuteranopia, or tritanopia, respectively. An additional rare case of CVD – monochromacy, occurs when neither photopigment nor only one type of photopigment is present [1].

Molecular investigations have revealed that the amino acid sequence of the L- and M-cone opsins shares 96% identity, only differing by seven amino acids in photopigment genes, while they exhibit a 46% identity with that of the S-cone opsins [3]. This particular genetic locus between the L- and M-cone opsins is prone to various cross-over events during the production of gametes in females [4]. Red-green CVD, which includes conditions such as protanopia and protanomaly (affecting L-cones), deuteranopia and deuteranomaly (affecting M-cones), constitutes the majority of color vision deficiencies[5,6]. Due to the scarcity of physiological references, tritanomaly and tritanopia, often considered acquired or autosomal-inherited [6], have not been included within the scope of this paper.

To simulate the color perception of color deficient observers, numerous recoloring methods have been developed over the years. Brettel et al. proposed a computerized simulation method of color appearance for dichromats, projecting the cone signals of the stimulus onto a reduced two-dimensional surface determined by a series of monochromatic anchor stimuli [7]. Yang et al. expanded the target group to anomalous trichromats, proposing a framework for computerized color simulation [8]. The cone fundamental of the abnormal cone was simulated by simply shifting it along the wavelength domain. Furthermore, Machado et al. proposed a two-step color appearance simulation method, involving transformations at the cone-signal stage and opponent-color space, applicable to both dichromats and anomalous trichromats [9]. A factor – 0.96 was adopted to scale the area ratio between L- and M-cone, to correct its inaccuracies by taking Brettel et al.'s model as the ground truth. Additionally, Machado et al. suggest applying an interpolation between the cone fundamentals of the L- and M-cone using the ratio of the peak wavelength shift to obtain an intermediate curve for the abnormal cone. However, the simulation results of Machado's method were criticized by many researchers due to its dependence on the choice of display. To solve the simulation inconsistency, Machado et al. subsequently published the RGB transformation matrices for CVD at various types and severities [10]. Addressing the overflow problem of RGB signals in simulated images, You and Park proposed a method to optimize the normalization step, ensuring the overall quality of recolored images for CVD [11].

Even though the models mentioned above take the cone space (LMS) as the base color space for the transformation from normal color vision to CVD, their underlying assumptions do not fully align with physiological realities. For example, when calculating the abnormal cone fundamentals, the aforementioned models directly manipulate the cone fundamentals of a standard observer through interpolation or peak wavelength shift, overlooking the fact that CVD arises from a shift in the peak wavelength of the absorptance spectrum[12]. In contrast, Yaguchi et al. proposed a physiological-based model for simulating the color appearance of anomalous trichromats based on the CIE 2006 physiological observer model [13]. In this model, the spectral quantal absorption of the abnormal L-/M-photopigment is assumed to shift along the wavenumber domain, considering that L- and M-photopigments exhibit similar shapes as a function of wavenumber [14–16]. Moreover, under the assumption that all observers perceive the same color for equal-energy-white (EEW), the simulated cone fundamentals for anomalous trichromats are normalized to EEW. The performance of Yaguchi et al.'s model has been validated through simulations of the Rayleigh match and a discrimination threshold experiment. Recently, Stockman and Rider fitted the photopigment absorbance spectra, as well as the macular and lens pigment density spectra, using 8th-order Fourier polynomial functions. This model can be extended to red-green color vision deficient observers by shifting the shape-invariant templates of photopigment spectra along the log wavelength scale [17].

Nonetheless, the difference in the spectral quantal absorption between L- and M-photopigments still exists. Consequently, for protanopes and deuteranopes with identical peak wavelengths in L- and M-photopigment, the cone fundamentals of L- and M-cones predicted by Yaguchi et al.'s model still do not overlap, in contrast to the theoretical results for dichromats. Additionally, the performance of Yaguchi et al.'s model does not align with our expectations based on our pilot examinations, particularly when simulating the pseudoisochromatic plate (seen in section 4.2). To bridge the gap between dichromats and anomalous trichromacy with different types and severities, this study proposes a new model for estimating the cone fundamentals of CVD by adjusting the shape of the absorption spectrum of the abnormal photopigment with its degree of abnormality. To verify the accuracy of this model, two validation experiments were conducted using the computerized pseudoisochromatic plate and Farnsworth Munsell 100 Hue test (FM-100) simulated using the proposed method, and the results were compared with those obtained from other simulation methods.

## 2. Methods

The simulation method for cone fundamentals is based on the CIE 2006 physiological observer model [18]. It assumes that individuals with color vision deficiency (CVD) have the absorptance spectra of abnormal photopigments replaced by a shifted one compared to individuals with normal color vision [19]. For instance, in protanomaly, the absorptance spectrum of the abnormal L-photopigment shifts towards the shorter wavelength region, approaching that of the normal M-photopigment. Similarly, in deuteranomaly, the absorptance spectrum of the abnormal M-photopigment shifts towards the longer wavelength region. For protanope, the abnormal L-photopigment has an identical absorptance spectrum as the normal M-photopigment; for deuteranope, the abnormal M-photopigment has an identical absorptance spectrum as the normal L-photopigment.

In the proposed model, the waveform of the abnormal photopigment gradually transitions between those of normal L- and M-photopigments based on its degree of abnormality.

Lamb and Baylor et al. have demonstrated that the L- and M-photopigments exhibit a similar shape on a log wavenumber scale [14,20], as measured by Dartnall in 1953, called Dartnall nomogram [21]. The invariance in the shape of photopigments has also been identified on both the log wavelength scale and the wavenumber scale [17].

Consequently, consistent with Yaguchi et al.'s model, this approach incorporates the shift in the absorptance spectrum along the wavenumber scale. Even though the spectral absorptance curves of L- and M-photopigments closely resemble each other [22,23], their slight differences can be amplified in the simulated cone fundamentals. Note that the spectral quantal absorptions of L- and M-photopigments have a peak wavenumber difference of 700 $cm^{-1}$, corresponding to a peak wavelength interval of 20 $nm$. To obtain the spectral quantal absorption of the abnormal photopigment, the first step is to shift that of the L- or M-photopigment along the wavenumber until its peak wavenumber matches that of the normal M- or L-photopigment, as expressed in Eqs. (1) and (2):

$$\log A_{Lo}(v) = \log A_L(v + \Delta v_o) \tag{1}$$

$$\log A_{Mo}(v) = \log A_M(v - \Delta v_o) \tag{2}$$

where $A_L(v)$ and $A_M(v)$ represent the low-density spectral quantal absorption of L- and M-photopigments for a normal trichromat, respectively; $A_{Lo}(v)$ and $A_{Mo}(v)$ denote the shifted $A_L(v)$ and $A_M(v)$ with the identical peak wavenumber as the normal M- or L-photopigment, respectively; $\Delta v_o$ is fixed at 700 $cm^{-1}$. Note that the absorptance spectra of three photopigments for a normal trichromat are the same as the 32-year 2° standard observer defined in the CIE publication [18]. The subsequent step involves modeling the continuous transition in the relative waveform of the abnormal cone photopigment between L- and M-photopigments, as illustrated below:

$$\log A'_{Lo} = \alpha \times \log A_{Lo} + (1-\alpha) \times \log A_{Mo} \tag{3}$$

$$\log A'_{Mo} = \alpha \times \log A_{Mo} + (1-\alpha) \times \log A_{Lo} \tag{4}$$

where $\alpha = (700 - \Delta v)/700$, $\Delta v$ is the peak wavenumber shift of the abnormal L- or M-photopigments absorption spectrum, indicating the degree of abnormality; $A'_{Lo}(v)$ and $A'_{Mo}(v)$ denote the reshaped low-density absorption spectra of the shifted L- and M-photopigments, respectively. The shape of the absorptance spectra gradually changes between that of the M-photopigment and the L-photopigment as the peak wavelength shifts. Afterwards, $A'_{Lo}(v)$ and $A'_{Mo}(v)$ should be shifted back to align with their corresponding peak wavenumbers, while maintaining their relative waveforms invariant. The absorption spectra of the abnormal L- and M-photopigments were obtained in Eqs. (5) and (6), respectively:

$$A'_L(v) = A'_{Lo}(v - \Delta v_o + \Delta v) \tag{5}$$

$$A'_M(v) = A'_{Mo}(v + \Delta v_o - \Delta v) \tag{6}$$

where $A'_L(v)$ refers to the low-density absorption spectrum of L-photopigment for protanomaly, and $A'_M(v)$ refers to the relative absorption spectrum of M-photopigment for deuteranomaly. The wavenumber shifts $\Delta v$ of 170, 350, 520, and 700 $cm^{-1}$ correspond to the wavelength shifts $\Delta \lambda$ of 5, 10, 15, and 20 $nm$, respectively. It is worth noting that although genetic studies have indicated that different genotypes of CVD correspond to distinct peak wavelength shifts for the L/M cone sensitivity curve [24], spectrally active polymorphisms may cause these peak wavelength shifts to fluctuate around the typical values. This implies the necessity of performing color simulations for CVDs with the peak wavelength continuously ranging between 0 and 20 $nm$. It is widely accepted that observers with a peak wavelength shift of L/M cone exceeding 8 $nm$ exhibit noticeable color vision abnormalities [25]. In the case of deuteranomaly, the $A'_M(\lambda)$ at different abnormality levels of peak wavelength shift $\Delta\lambda$ have been depicted in Fig. 1. Note that the shape of $A'_L(\lambda)$ is identical to that of $A'_M(\lambda)$ with complementary $\Delta\lambda$ ($\Delta\lambda_L + \Delta\lambda_M = 20$ $nm$). It can be observed that when $\Delta\lambda = 20$ $nm$, $A'_M(\lambda)$ is identical to $A_L(\lambda)$. Regarding the self-screening effect, the obtained low-density absorption spectra of three photopigments are used to calculate the absorptance spectra of three cones in terms of quanta, namely, $\alpha_{i,l}(\lambda)$, $\alpha_{i,m}(\lambda)$, $\alpha_{i,s}(\lambda)$, following the CIE170-1 approach[18], as given in Eqs. (7) - (9):

$$\alpha_{i,l}(\lambda) = 1 - 10^{[-D_{\tau,\max(L-cones)} A_{i,0(L-pigment)}(\lambda)]} \tag{7}$$

$$\alpha_{i,m}(\lambda) = 1 - 10^{[-D_{\tau,\max(M-cones)} A_{i,0(M-pigment)}(\lambda)]} \tag{8}$$

$$\alpha_{i,s}(\lambda) = 1 - 10^{[-D_{\tau,\max(S-cones)} A_{i,0(S-pigment)}(\lambda)]} \tag{9}$$

where $D_{\tau,max}$ refers to the peak optical density of each photopigment, which is a constant related only to the field size $f_s$; $A_{i,0}(\lambda)$ of each photopigment refers to the low-density absorption spectra normalized at the peak optical density. When $f_s = 2°$, $D_{\tau,max} = 0.5, 0.5, 0.4$ for the L-, M-, S-photopigments, respectively; when $f_s = 10°$, $D_{\tau,max} = 0.38, 0.38, 0.3$ for the L-, M-, S-photopigments, respectively. According to the CIE 2006 physiological observer model, the cone fundamental sensitivity functions in terms of quanta, $\bar{l}_q(\lambda)$, $\bar{m}_q(\lambda)$, $\bar{s}_q(\lambda)$, are estimated by multiplying the quantal spectral absorption by the transmittance of ocular media $\tau_{ocul}(\lambda)$, macular pigment $\tau_{macula}(\lambda)$, as calculated in Eqs. (10) - (12):

$$\bar{l}_q(\lambda) = \alpha_{i,l}(\lambda) \cdot \tau_{macula}(\lambda) \cdot \tau_{ocul}(\lambda) \tag{10}$$

$$\bar{m}_q(\lambda) = \alpha_{i,m}(\lambda) \cdot \tau_{macula}(\lambda) \cdot \tau_{ocul}(\lambda) \tag{11}$$

$$\bar{s}_q(\lambda) = \alpha_{i,s}(\lambda) \cdot \tau_{macula}(\lambda) \cdot \tau_{ocul}(\lambda) \tag{12}$$

Note that the $\tau_{ocul}(\lambda)$ systematically changes with the observer's age, and $\tau_{macula}(\lambda)$ is related to the field size $f_s$. The peak optical density of the macular pigment $D_{\tau,max,macula}$ has the value of 0.35 and 0.095 at the 2° and 10° field of view, respectively. The quantal cone fundamentals $\bar{l}_q(\lambda)$, $\bar{m}_q(\lambda)$, $\bar{s}_q(\lambda)$ can be converted into cone fundamentals in terms of energy $\bar{l}(\lambda)$, $\bar{m}(\lambda)$, $\bar{s}(\lambda)$ by multiplying $\lambda$, and then renormalizing to the maximum values. As we assume that all individuals perceive the same color for EEW, $\bar{l}(\lambda)$, $\bar{m}(\lambda)$, $\bar{s}(\lambda)$ are normalized to EEW. In other words, the cone responses of the EEW spectrum for an individual observer with CVD are the same as those of the standard observer. EEW is considered the anchor for renormalization. Thus, the cone fundamentals of the abnormal L or M cone for CVD should be renormalized, as shown below:

$$\bar{l}_a'(\lambda) = (L_{EEW} / L'_{EEW})\bar{l}'(\lambda) \tag{13}$$

$$\bar{m}_a'(\lambda) = (M_{EEW} / M'_{EEW})\bar{m}'(\lambda) \tag{14}$$

where $\bar{l}'(\lambda)$, $\bar{m}'(\lambda)$ are cone fundamentals of the abnormal L- and M-cone in terms of energy obtained from the previous step for an individual observer with CVD; $L_{EEW}$ and $M_{EEW}$ are the cone signals of EEW for the CIE 2006 32-year standard observer with normal color vision, while $L'_{EEW}$ and $M'_{EEW}$ are those for an individual observer with CVD; $\bar{l}_a'(\lambda)$, $\bar{m}_a'(\lambda)$ are the normalized cone fundamentals of the abnormal L- and M-cone, replacing the $\bar{l}'(\lambda)$, $\bar{m}'(\lambda)$ as the generated cone fundamentals for CVD. Figures 2(a) and 2(b) depict the cone fundamentals generated by the proposed method for protanomaly (or protanope) and deuteranomaly (or deuteranope) with varying peak wavelength shifts, respectively. Note that in the case of protanope (with $\Delta\lambda = 20\ nm$), the abnormal L-cone exhibits an identical relative waveform to the normal M-cone but with distinct peak sensitivity. This is to ensure that the covered area of the abnormal L-cone is the same as that of the normal L-cone, maintaining consistent cone responses for EEW. A similar phenomenon is observed for deuteranope, as shown in Fig. 1(b).

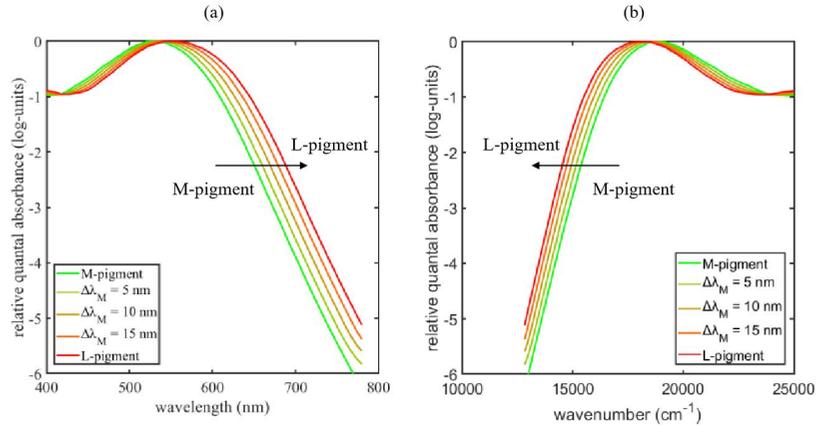

Fig. 1. The absorptance spectra of L- and M-photopigments for deuteranomaly, featuring $\Delta\lambda$ of 5, 10, 15, and 20 $nm$ represented by different colors. (a). Along the wavelength scale (b). Along the wavenumber scale

Subsequently, the cone fundamentals produced for individuals with CVD can be converted to the $\bar{x}_F(\lambda)$, $\bar{y}_F(\lambda)$, $\bar{z}_F(\lambda)$ color matching functions (CMFs) using the transformation matrix defined in CIE170-2. It is noteworthy that the transformation matrices for the 2° and 10° stimuli are distinct. In cases where the field of view falls between 2° and 10°, the transformation matrix can be acquired through interpolation between the two matrices. Then the $X_F Y_F Z_F$ tristimulus can be calculated from the radiance spectrum using the $\bar{x}_F(\lambda)$, $\bar{y}_F(\lambda)$, $\bar{z}_F(\lambda)$ CMFs.

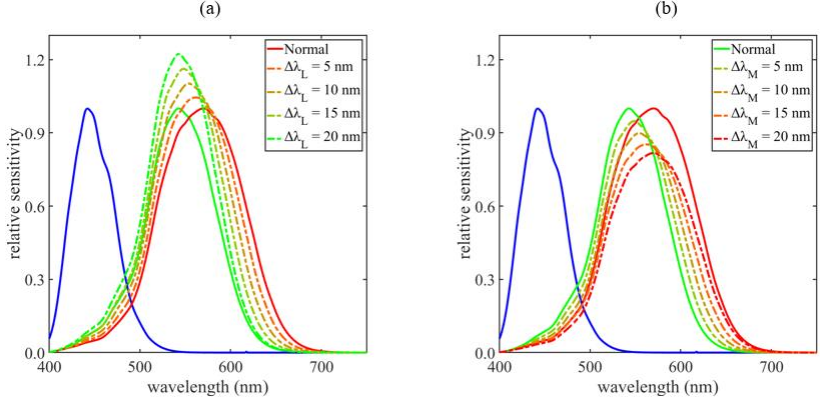

Fig. 2. (a). The 2° normalized cone fundamentals for protanomaly with 5, 10, 15, 20 *nm* peak wavelength shift. (b). The 2° normalized cone fundamentals for deuteranomaly with 5, 10, 15, 20 *nm* peak wavelength shift.

It is worth noting that the photopigment absorption spectrum template in Eqs. (1), (3), and (5) can be adjusted to account for the polymorphism of the L cone—serine (L(ser180)) or alanine (L(ala180)) at position 180 of the L photopigment opsin gene—by substituting the standard template with the L(ser180)/L(ala180) template. The optical density of the lens, macular and three photopigments can also been adjusted in the physiological model to generate individual cone fundamentals for CVD. However, since the primary objective of this paper is to develop a universal template for photopigment absorptance in CVD observers with varying types and degrees of abnormality, the other parameters (except for $\Delta\lambda_L$ or $\Delta\lambda_M$) in the physiological observer model are kept consistent with those of the CIE standard observer to validate the proposed template.

## 3. Color Appearance simulation

To assess the efficacy of the proposed model in reproducing images for individuals with CVD, the model was applied to simulate test images, accounting for variations in the type and severity of CVD. Applying this model to digital images requires considerations of the spectral information corresponding to the RGB values of each pixel. Subsequently, the $X_F Y_F Z_F$ tristimulus for CVD can be computed using the simulated $\bar{x}_F(\lambda)$, $\bar{y}_F(\lambda)$, $\bar{z}_F(\lambda)$ and the display spectrum. The RGB values for the simulated images in CVD are then derived from $X_F Y_F Z_F$ tristimulus considering the properties of the display. Therefore, it is necessary to characterize the display, involving obtaining the spectra of the three primaries and their respective luminance curves. A 27-inch Dell LCD monitor (1920 × 1080 pixels) was utilized to display both the original and simulated images in this study. Due to the channel independence and chromaticity constancy, the display was characterized using the classical gamma-offset-gain (GOG) model where the parameters of the luminance curve for each channel and its chromaticity were determined from the measured XYZ tristimulus values of the training dataset. This characterization model was then used to estimate the luminance of each channel for a given set of (R, G, B) drive values. Owing to the channel independence of the display, the corresponding spectrum for the specified (R, G, B) set can be determined as the sum of the spectra from the three channels, as illustrated below:

$$spd_R = \frac{L_R - L_{dark}}{L_{R,max} - L_{dark}} \cdot \left(spd_{R,max} - spd_{dark}\right) \quad (15)$$

$$spd_G = \frac{L_G - L_{dark}}{L_{G,max} - L_{dark}} \cdot \left(spd_{G,max} - spd_{dark}\right) \quad (16)$$

$$spd_B = \frac{L_B - L_{dark}}{L_{B,max} - L_{dark}} \cdot \left(spd_{B,max} - spd_{dark}\right) \quad (17)$$

$$spd_{RGB} = spd_R + spd_G + spd_B + spd_{dark} \quad (18)$$

Where $L_R$, $L_G$, $L_B$ are the luminance of the R, G, B channel, respectively; $L_{R,max}$, $L_{G,max}$, $L_{B,max}$ are the maximum luminance of the R, G, B channel, respectively; $spd_{R,max}$, $spd_{G,max}$, $spd_{B,max}$ are the spectrum at the maximum luminance of the R, G, B channel, respectively; $spd_{dark}$ and $L_{dark}$ refer to the measured spectrum and the corresponding luminance at the dark point; $spd_{RGB}$ refers to the display spectrum for the specified (R, G, B) set. The spectrum estimated using the method outlined in Eqs. (15) – (18) closely aligns with the measured spectrum obtained using a Konica Minolta CS-2000 spectroradiometer. The minor discrepancies observed confirm the accuracy of the display characterization model, with a maximum chromaticity error of less than 0.0030 in the $u_F'v_F'$ diagram, a level deemed imperceptible [26].

Figure 3 depicts simulations of the test image 'fruits' at various degrees of severity, identified by the peak wavelength shift, using the proposed model for protanomaly and deuteranomaly (based on the CIE 2006 2° LMS). As the severity of CVD increases, the color contrast in the simulated image gradually diminishes. Consequently, the color appearance differences between reddish and greenish objects decrease, in accordance with real-world observations. Additionally, it is notable that the reddish objects appear darker for protanomaly (or protanope) compared to deuteranomaly (or deuteranope). The ($u_F'$, $v_F'$) chromaticity coordinates of four simulated colors in the test image for anomalous trichromats at different degrees of severity have been presented in Figs. 4(a) and 4(b), corresponding to protanomaly and deuteranomaly, respectively. The four selected colors include red from tomato, green from broccoli, yellow from paprika, purple from grape. Remarkably, for each color stimulus, the coordinates with different degrees of abnormality (including normal trichromacy and dichromacy), as simulated, consistently lie on the confusion line. For either protanope or deuteranope, all the confusion lines converge to a confusion point [27], consistent with the previous findings [28]. Furthermore, the color coordinates exhibit a uniform and proportional shift as the degree of color vision abnormality increases. Additionally, it can be observed that the chromaticity coordinates shift towards that of EEW with the increasing degree of abnormality.

## 4. Experimental validation

To further validate the effectiveness of the simulation results using the proposed model, we simulated the perceived colors for individuals with CVD in two color deficiency examination methods: the FM-100 test and the pseudoisochromatic plate test. Observers with normal color vision then participated in these tests using the simulated versions of CVD, to resemble the color vision tests for individuals with CVD. If the results using the simulated test are consistent with those of the original test for CVD, the effectiveness of the proposed model can be confirmed. Furthermore, we compared the simulation results with those obtained from other CVD simulation models to further assess the performance and reliability of the proposed model. All colorimetric calculations were performed under a 2° viewing condition due to the small size of the target stimuli in each complex image.

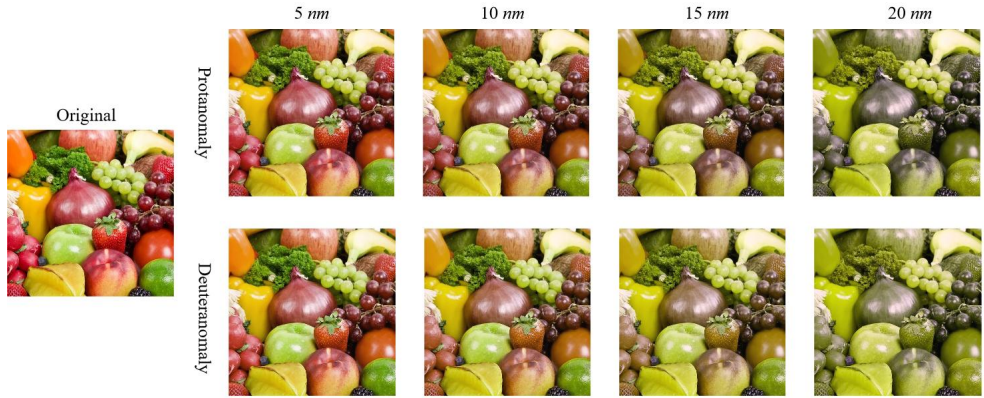

Fig. 3. Simulations of digital image 'fruits' for anomalous trichromacy with the peak wavelength shifts of 5, 10, 15, 20 $nm$, obtained using the proposed model.

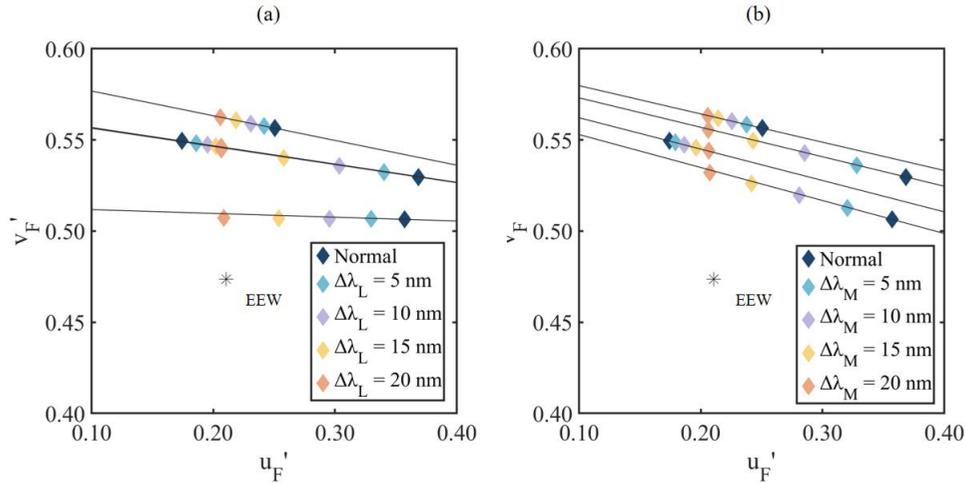

Fig. 4. The $u_F'v_F'$ chromaticity coordinates of four simulated colors selected from the test image 'fruit' for individuals with normal color vision and CVD (with different degrees of abnormality). (a) Protanomaly. (b) Deuteranomaly.

*4.1 FM-100 test*

4.1.1 Method

The FM-100 test comprises 85 caps distributed across four rows, as presented in Fig. 5(a). The first row contains 22 caps, while each of the remaining rows consists of 21 caps. Each cap is assigned a unique number ranging from 1 to 85. During the testing procedure, participants are instructed to sort the caps in each row until they observe a gradient and subsequent hue change. The total error scores (TES), computed from the sorting results, serve as a metric for assessing an individual's color discrimination ability, with a range of 0-16 considered 'superior', 16-100 classified as 'average', and a score exceeding 100 as 'low' color discrimination ability. Color-vision-deficient observers always show significantly higher TES scores compared to those with normal color vision. Additionally, a polar map can be generated to visually illustrate the distribution of error scores and categorize different types of CVD [29]. Specifically, protanomaly/protanopia tends to make more errors around caps of No. 19 ± 5 and 65 ± 3, while deuteranomaly/deuteranopia exhibits a higher frequency of errors around caps of No. 15 ± 4 and 59 ± 3 [30], as shown in Fig. 5(d).

Previous research has demonstrated the agreement in the TES between the computerized and original FM-100 tests [31]. To conveniently simulate the color appearance of caps for CVDs, we developed a computerized version of the FM-100 test, wherein accurately reproduced the color appearance of each cap on the display. The radiance spectrum of each sample was measured under illumination with D65 chromaticity using a Konica Minolta CS-2000 spectroradiometer, employing a 0/45 viewing geometry. The D65 illumination was generated by a multi-channel LED lighting booth, providing a continuous spectrum across the visible wavelength range. Note that we reproduced the CIE 1976 $L^*a^*b^*$ values of each cap observed under the D65 illumination on the display, instead of using $X_F Y_F Z_F$ tristimulus. The reproduction errors were found to be below 2 $\Delta E_{ab}^*$ compared to the $L^*a^*b^*$ values of the physical cap, indicating a high accuracy in color appearance reproduction on the display [26]. For normal color vision, the $X_F Y_F Z_F$ tristimulus of each cap was calculated using the CIE 2006 32-year 2° CMFs. The perceived color appearance of all 85 caps was simulated for protanomaly and deuteranomaly with the peak wavelength shift of 18 $nm$ (a situation of serious abnormality), as depicted in Figs. 5(b) and 5(c), respectively. The CMFs used to compute the $X_F Y_F Z_F$ tristimulus were derived from the method proposed in section 2.

Five observers with CVD (D1 ~ D5) and five observers with normal color vision (N1 ~ N5), as diagnosed by the Ishihara test, participated in the experiment, with ages ranging between 23 and 26 years. All participants completed the physical FM-100 test under D65 illumination provided by a Datacolor Tru-Vue lighting booth. Additionally, the participants with CVD were assigned the computerized FM-100 test, involving sorting colors reproduced for normal color vision, as illustrated in Fig. 5(a). In contrast, the two participants with normal color vision were tasked with sorting colors simulated for both protanomaly and deuteranomaly ($\Delta\lambda$ = 18 $nm$). During the experiment, the 85 caps were presented row by row on a black screen, replicating the conditions of the physical FM-100 test. Each cap was rendered in an elliptical shape, providing a realistic viewing experience in terms of shape and dimensions. For each row, all the caps except those at the polar ends were randomly displayed and their positions can be manually manipulated by the observers using the mouse. Participants were instructed to arrange the displayed colors in a gradient order. The total duration of the experiment and the sorting sequence were recorded for subsequent analysis. All experimental procedures were approved by Beijing Institute of Technology Institutional Review Board.

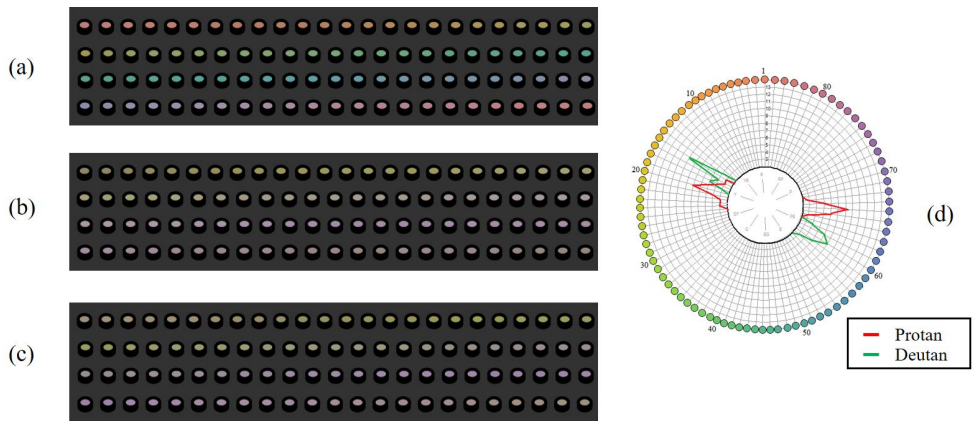

Fig. 5. Layout of the computerized FM-100 test featuring 83 caps arranged in four rows. (a). Color reproduction for normal color vision. (b). Color reproduction for protanomaly with $\Delta\lambda_L$ of 18 $nm$. (c). Color reproduction for deuteranomaly with $\Delta\lambda_M$ of 18 $nm$. (d). Example of a polar map visualizing the error distribution for typical protanomaly and deuteranomaly.

4.1.2 Results

Table 1 summarizes the TES and completion duration when performing both the physical and computerized FM-100 tests conducted by ten observers (D1 ~ D5, N1 ~ N5). The results from the physical FM-100 test demonstrate that individuals with normal color vision efficiently perform the task with perfect sorting, while individuals with CVD exhibit higher TES and longer completion durations. In the computerized test, the TES and duration of CVD individuals (D1 ~ D5) show minor changes compared to the physical test, indicating the consistency between the physical and computerized tests. In the computerized test for CVD individuals (N1 ~ N5) viewing simulated colors of CVD, most of them consistently received a diagnosis of low color discrimination (> 100). Their TES values were significantly higher than the physical test results, indicating a reduced color discrimination ability for the simulated CVD.

Additionally, Fig. 6 presents the polar maps for both physical and computerized FM-100 tests. As illustrated in Figs. 6(a) – 6(d), two color-deficient observers (D1 and D2) are consistently classified as deuteranomaly by both physical and computerized tests, as the error distribution aligns with the permutation characteristics typically observed for deuteranomaly. This suggests that the computerized test yields similar examination results, but higher TES compared to that of the physical test. In the case of normal observers (N1, N2) viewing the simulated protanomaly vision, the polar maps exhibit a significant dispersion of errors between No. 50 and No. 85, centered around the confusion line of protan-type CVD. When N1 and N2 view the simulated deuteranomaly vision, the sorting errors exhibit clustering characteristics along the confusion line of deutan-type CVD in the polar map. The consistency in the results of the FM-100 test between color-deficient observers viewing the normal test and those with normal color vision viewing the simulated test for CVD preliminarily indicates that the proposed model can effectively reproduce the color appearance of CVD.

Table 1. A summary of the experiment results, including the TES and completion duration, for the physical and computerized FM-100 test

| Observers | Physical Test | | | Computerized Test | | | |
|---|---|---|---|---|---|---|---|
| | TES | Classification | Duration (seconds) | Simulation | TES | Classification | Duration (seconds) |
| D1 | 88 | Average | 399 | Normal | 256 | Low | 373 |
| D2 | 140 | Low | 679 | Normal | 272 | Low | 566 |
| D3 | 48 | Average | 2420 | Normal | 92 | Average | 2590 |
| D4 | 84 | Average | 914 | Normal | 168 | Low | 1134 |
| D5 | 112 | Low | 836 | Normal | 228 | Low | 595 |
| N1 | 0 | Superior | 875 | Protan | 68 | Average | 2269 |
| | | | | Deutan | 112 | Low | 1437 |
| N2 | 16 | Superior | 629 | Protan | 236 | Low | 1309 |
| | | | | Deutan | 216 | Low | 1061 |
| N3 | 0 | Superior | 440 | Protan | 152 | Low | 706 |
| | | | | Deutan | 124 | Low | 752 |
| N4 | 8 | Superior | 621 | Protan | 40 | Average | 745 |
| | | | | Deutan | 60 | Average | 652 |
| N5 | 16 | Superior | 946 | Protan | 164 | Low | 1391 |
| | | | | Deutan | 132 | Low | 1129 |

*4.2 Pseudoisochromatic plates*

### 4.2.1 Method

To assess the accuracy of simulated color appearances for individuals with CVD and conduct a comparative analysis of different CVD simulation methods, four pseudoisochromatic plates have been selected from the Waggoner Computerized Color Vision test (WCCVT) – an advanced color vision testing suite developed by ophthalmic professionals. These plates include two groups of vanishing plates, as shown in Figs. 7(a) and 7(b), and two groups of classification plates used for detecting protanomaly/protanope (see Fig. 7(c)) and deuteranomaly/deuteranope (see Fig. 7(d)), respectively. Additionally, a set of plates has been generated for each group of test plates, with varying numbers but consistent color combinations of the foreground numeral and background There are 7 plates in each vanishing group, and 13 plates in each classification group. Note that these numbers could be either single or double digits. This approach was implemented to ensure the randomization of content numbers across the plates utilized in the subsequent experiment, thereby minimizing bias in numeral recognition gained from the prior experience.

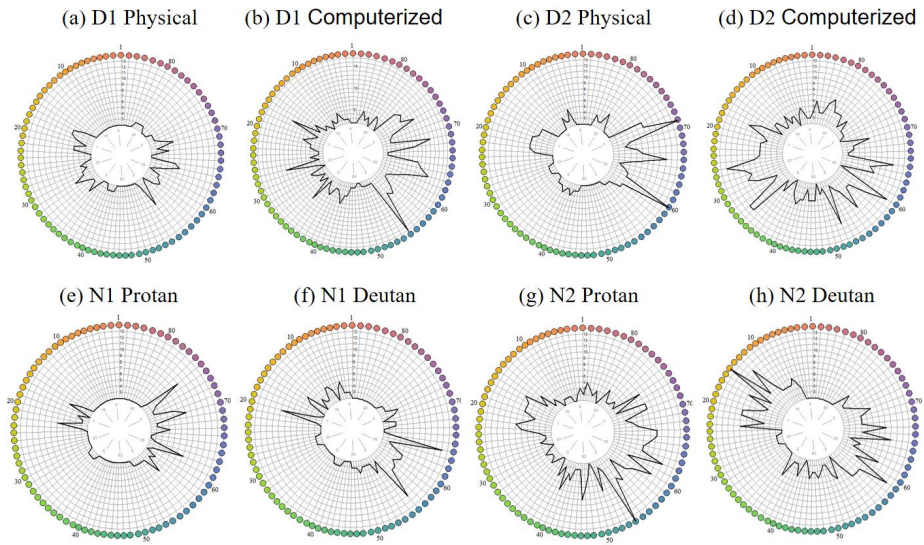

Fig. 6. Typical polar maps of physical and computerized FM-100 tests for ten observers. (a). Physical test result for D1. (b). Computerized test result for D1. (c). Physical test result for D2. (d). Computerized test result for D2. (e). The result of computerized test with simulating protanomaly vision for N1. (f). The result of computerized test with simulating deuteranomaly vision for N1. (g). The result of computerized test with simulating protanomaly vision for N2. (h). The result of computerized test with simulating deuteranomaly vision for N2.

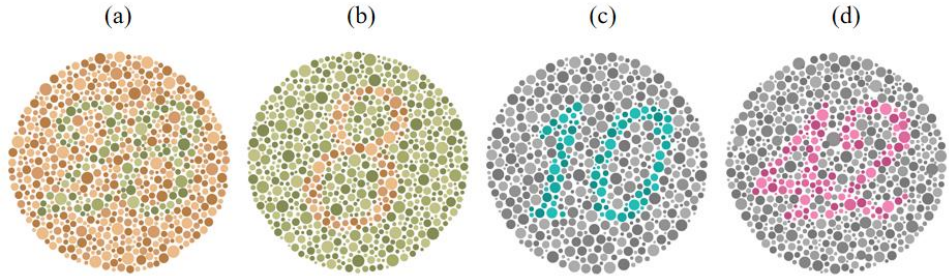

Fig. 7. The four pseudoisochromatic plates chosen from the Waggoner Computerized Color Vision test for color appearance simulation. (a-b). Vanishing plates. (c). Classification plate for protanomaly/protanope. (d) Classification plate for deuteranomaly/deuteranope.

For each plate, five distinct methods were employed to simulate the four groups of computerized pseudoisochromatic plates as perceived by color-deficient observers with variations in CVD type and abnormality. These methods include Asano's method (which involves the absorbance spectra shifting in the wavelength domain), Yaguchi's method (which involves absorbance spectra shifting in the wavenumber domain), Machado's method (which utilizes direct interpolation between the cone fundamentals of the L- and M-cone), Stockman's method (which involves the absorbance spectra shifting in the log wavelength domain) and the method proposed in this paper. Note that for Machado's method, the transformation matrices for CVD with different peak wavelength shifts have been adopted for simulation. These matrices accommodate displays featuring the sRGB color gamut, consistent with the Dell display employed in this study. Two distinct types of CVD, protanomaly/protanope and deuteranomaly/deuteranope, with eight levels of abnormality ranging between 0.5 and 1 ($\alpha$ = 0.5, 0.65, 0.75, 0.8, 0.85, 0.9, 0.95, 1) were simulated for each computerized test plate using the four methods. The Dell LCD monitor mentioned above has been used for presenting the simulated images. It is worth noting that in the Stockman's method, considering that the L(ala180) template can be derived by shifting L(ser180) template to 0.002125 log10 $nm$ towards shorter wavelength, when simulating CVD, the L-cone adopts the photopigment absorbance template of L(ser180) ($\lambda_{max}$ = 551.9 $nm$), with a 23.7 nm interval apart from the M-cone ($\lambda_{max}$ = 529.8 $nm$). In contrast, the CIE 2006 32-year 2° standard observer has this interval at 20 nm. To ensure consistency when comparing with Stockman's model, we use the same measure of color vision abnormality, $\alpha_\lambda$, defined as the ratio of $\Delta\lambda_L$ or $\Delta\lambda_M$ to the $\lambda_{max}$ interval between the standard L- and M-photopigment. This value ranges from 0 to 1, with higher $\alpha_\lambda$ values indicating more severe CVD.

Figures 8(a) and 8(b) present an example of the protan-classification plate (Fig. 7(c)) simulated by five methods at three levels of abnormality ($\alpha_\lambda$ = 0.75, 0.9, 1) for protanomaly/protanope and deuteranomaly/deuteranope, respectively. It can be observed that our model stood out in CVD classification by presenting indistinguishable and discernible numerals for the protanomaly/protanope and deuteranomalous/deuteranope, respectively. In other words, protanomaly/protanope experiences difficulty in identifying the numerals, whereas deuteranomalous/deuteranope can still easily recognize the numerals, aligning with the expected performance of two CVD types for the protan-classification plate. For the other four methods, the numerals remain discernible regardless of the CVD type, indicating a failure in the classification of CVD type. Additionally, it is worthwhile to highlight that the grey dots on the simulated plates using the proposed method undergo a slight change toward a reddish direction with the severity of CVD. A similar trend has also been observed in the simulation using Yaguchi's method. It could be attributed to the 'EEW normalization' of the abnormal cone fundamental (see Eqs. (13) and (14)) in both methods. This assumption ensures that the perceived color appearance of EEW remains constant across all individuals. However, it's important to note that the white point of the Dell display is D65. As a result, the perceived color of the display's white point varies depending on the type and severity of CVD. The simulated achromatic dots for individuals with CVD, which tend to shift towards the chromaticity of EEW, exhibit a slight reddish appearance compared to D65, as shown in Fig. 8.

The experimental evaluation involves four groups of pseudoisochromatic plates. Through this approach, our goal is to validate the results generated by our model and to compare the visual performance of the five CVD simulation methods in color appearance simulation. A total of 320 test plates (4 groups × 2 types × 8 levels of abnormality × 5 methods) were generated for each observer, each containing a randomly assigned numeral. During the experiment, each observer sat in front of the monitor with a horizontal viewing distance of approximately 45 $cm$, in a dark room. The task for each observer is to identify the numeral presented on the plate. The test plates were randomly presented, with each image displayed for 3 $s$. After the image disappeared, participants were given an additional 8 $s$ to respond by pressing the number they perceived. If they did not see any number, they were instructed to press the space bar. Each test plate was displayed twice during the experiment. Note that the white background against the test plate has been adjusted with the type and severity of CVD, as well as the simulation method, to maintain color consistency across the entire display.

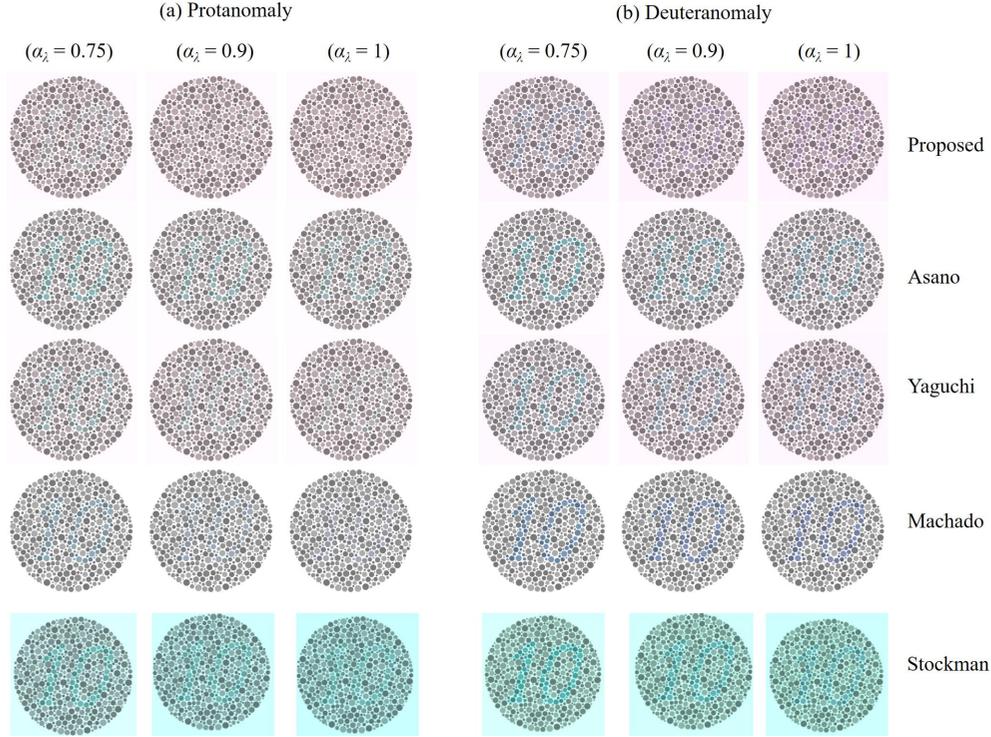

Fig. 8. The simulations of the protan-classification plate (Fig. 7(c)) using five methods at three levels of abnormality ($\alpha_\lambda$ = 0.75, 0.9, 1) for two types of CVD: (a). protanomaly/protanope, (b). deuteranomaly/deuteranope.

Ten observers with normal color vision, as diagnosed by the Ishihara test, participated in this experiment. Their ages range between 21 to 26 years with an average of 22.9 and a standard deviation of 1.37. All experimental procedures were approved by Beijing Institute of Technology Institutional Review Board.

4.2.2 Results

For each plate, the recognition rate was calculated as the ratio of correct answers to the total number of responses from ten observers across two repetitions. The relationship between the recognition rate and the peak wavelength shift $\Delta\lambda$ has been fitted by a psychometric function. Figure 9 compares the curves of recognition rate against $\alpha_\lambda$ among five simulation methods for five groups of test plates and two types of CVD. A higher recognition rate indicates that the number is easier to identify. The $\Delta\lambda$ corresponding to the $\alpha_\lambda$, at which the recognition rate of 50%, is considered the diagnosis threshold for the specific type of CVD.

Figure 9 illustrates a strong correlation between the recognition rate and the peak wavelength shift. Regardless of the type of test plate, it is evident that the plates simulated using Asano's, Yaguchi's, Stockman's methods consistently exhibit recognition rates higher than 50%, even under the most severe CVD condition ($\alpha_\lambda$ = 1). Notably, the recognition rate of the test plate simulated by Asano's method remains consistently close to 100%. This indicates that all observers were able to accurately recognize the numerals on plates simulated by both Asano's, Yaguchi's, Stockman's methods. This finding suggests that when employing

Asano's, Yaguchi's, Stockman's methods, the color difference between the numeral and background in the simulated test plate is larger than that perceived by individuals with CVD in the original test plate. Additionally, as Stockman's method does not include the white point normalization, the simulated test plates appear blue-greenish, in contrast to the near-neutral backgrounds seen in other simulation methods.

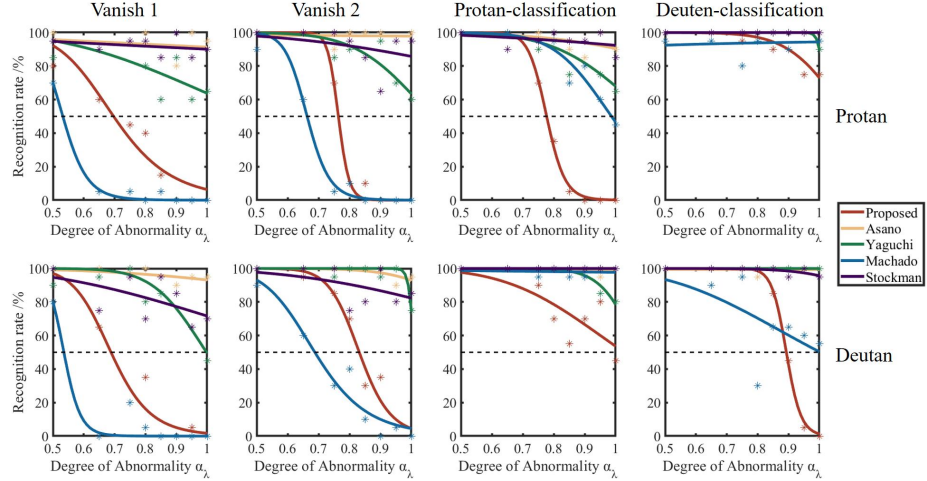

Fig. 9. Psychometric functions fitting the relationship of the recognition rate against $\alpha_\lambda$. Five curves in each subfigure correspond to five simulation methods. Four columns correspond to four groups of test plate. Two rows correspond to two different types of CVD (protan, deutan).

For Machado's method, the recognition thresholds $\Delta\lambda$ of two vanishing plates are (10.6 *nm*, 13.3 *nm*) and (10.7 *nm*, 13.8 *nm*) for the protan-type and deutan-type CVD, respectively. For the proposed method, the recognition thresholds $\Delta\lambda$ for vanishing plates are (13.9 *nm*, 15.3 *nm*) and (13.8 *nm*, 16.6 *nm*) for the protan-type and deutan-type CVD, revealing a substantially higher threshold compared to Machado's method. However, the two methods exhibit different performances for classification plate groups. The numeral in the classification plate simulated by Machado's method is always visible, with a recognition rate consistently higher than 50%. In contrast, as simulated by the proposed method, the protan-classification plate has the numeral visible for deuteranomaly/deuteranope, but not for protanomaly/protanope. The opposite results can be observed for the deutan-classification plate. This finding indicates that the visual performance of the simulated plates using the proposed method closely aligns with the actual detection outcomes, showcasing its effective classification capabilities that set it apart from the other three methods.

## 5. Conclusion and future work

We proposed a physiological-based color appearance simulation model for CVD, incorporating the waveform change of the absorptance spectrum with the peak wavelength shift. This model unifies the anomalous trichromacy and dichromacy within a comprehensive framework. Further verification experiments of the FM-100 test and pseudoisochromatic plates demonstrated the high consistency of the proposed model in color perception with color-vision-deficient individuals. Furthermore, a comparative analysis has been implemented across various well-known models, revealing that the proposed method outperforms them in terms of visual consistency with the performance of CVD.

Regarding individual variations in color perception, the perceived color appearance of individual observers is influenced not only by the peak wavelength of the photopigment absorptance spectrum but also by other physiological parameters of the human visual system, such as the optical density of the lens, macula, and three photopigments [32]. Further discussions on the individual differences in cone fundamentals will be conducted in the subsequent sections of this work.

Moreover, the proposed model offers the advantage of digital image processing, rendering it highly convenient for applications in displays. In the future, we will further investigate the influence of the display primaries and the 'normalization anchor' on the simulated color appearance for CVD, and optimize the simulation method to adapt to displays varying in color gamut and white point.

**Funding.** This project is financially supported by National Key Research and Development Program of China (2023YFB3611500), National Natural Science Foundation of China (62205018, 62475013, 62332003).

**Disclosures.** The authors declare no conflicts of interest.

**Data Availability.** Data underlying the results presented in this paper are not publicly available at this time but may be obtained from the authors upon reasonable request.